\documentclass[journal]{IEEEtran}

\usepackage[utf8]{inputenc}

\usepackage{psfrag,epsfig,graphics}
\usepackage{amsmath,amsthm,amssymb,multirow}
\usepackage{mathbbol}
\usepackage{amssymb}             %

\DeclareSymbolFontAlphabet{\amsmathbb}{AMSb}%
\usepackage{mathabx} %

\usepackage{booktabs}
\usepackage{graphicx,amsmath,graphicx}
\usepackage{caption}
\usepackage[position=b]{subcaption}

\usepackage[noadjust]{cite}
\usepackage{multirow}
\usepackage[lined,linesnumbered,ruled]{algorithm2e}

\newcommand{\cb}[1]{{\boldsymbol{#1}}}
\newcommand{\cp}[1]{\ifmmode {\mathcal{#1}}\else ${\mathcal{#1}}$\fi}

\newcommand{\bA}{\boldsymbol{A}}

\newcommand{\bM}{\boldsymbol{M}}

\newcommand{\bQ}{\boldsymbol{Q}}

\newcommand{\bW}{\boldsymbol{W}}
\newcommand{\bX}{\boldsymbol{X}}
\newcommand{\bY}{\boldsymbol{Y}}

\newcommand{\ba}{\boldsymbol{a}}

\newcommand{\bm}{\boldsymbol{m}}

\newcommand{\bh}{\boldsymbol{h}}

\newcommand{\be}{\boldsymbol{e}}
\newcommand{\bg}{\boldsymbol{g}}

\newcommand{\by}{\boldsymbol{y}}

\newcommand{\bx}{\boldsymbol{x}}

\usepackage[euler]{textgreek}
\usepackage[colorinlistoftodos]{todonotes}

\newcommand{\balpha}{\boldsymbol{\alpha}}

\newcommand{\bomega}{\boldsymbol{\omega}}

\newcommand{\bTheta}{\boldsymbol{\Theta}}

\newcommand{\Ex}{\amsmathbb{E}}

\newcommand{\diag}{\operatorname{diag}}

\usepackage{color}  %
\definecolor{red}{rgb}{0.0, 0.0, 0.0}
\def\cred{\textcolor{red}}

\definecolor{darkgreen}{rgb}{0., 0.4, 0.}
\definecolor{amber}{rgb}{1.0, 0.49, 0.0}
\definecolor{orange}{rgb}{1.0, 0.4, 0.0}

\usepackage[usestackEOL]{stackengine}
\newcommand{\bigtriangleq}{\mathbin{\setstackgap{S}{0pt}\stackMath\Shortstack{\smalltriangleup\\ =}}}

\title{
Model-Based Deep Autoencoder Networks for\\ Nonlinear Hyperspectral Unmixing
}

\author{Haoqing Li, Ricardo A. Borsoi, Tales Imbiriba, Pau Closas, Jos\'e C. M Bermudez, Deniz Erdo{\u{g}}mu{\c{s}}
\thanks{This work has been supported by the National Science Foundation under Awards CNS-1815349 and ECCS-1845833, and the National Council for Scientific and Technological Development (CNPq) - Grants Nos. 304250/2017-1 and 409044/2018-0.
H. Li, T. Imbiriba, P. Closas and D. Erdo{\u{g}}mu{\c{s}} are with the ECE Dept., Northeastern University, Boston, MA 02115, USA (e-mail:
\mbox{li.haoq, t.imbiriba, closas, d.erdogmus, \emph{user}@northeastern.edu})
R.A. Borsoi and J.C.M. Bermudez are with the EEL--UFSC, Florian\'opolis 88040-370, SC, Brazil.
R.A. Borsoi is also with the Lagrange Laboratory (CNRS, OCA), Universit\'e  C\^ote  d'Azur, Nice, France. (e-mail: \mbox{raborsoi@gmail.com}).

}

}

\begin{document}

\maketitle

\begin{abstract}
Autoencoder (AEC) networks have recently emerged as a promising approach to perform unsupervised hyperspectral unmixing (HU) by associating the latent representations with the abundances, the decoder with the mixing model and the encoder with its inverse. AECs are especially appealing for nonlinear HU since they lead to unsupervised and model-free algorithms. However, existing approaches fail to explore the fact that the encoder should invert the mixing process, \cred{which might} reduce their robustness. In this paper, we propose a model-based AEC for nonlinear HU by considering the mixing model \cred{a nonlinear fluctuation} over a linear mixture. Differently from previous works, we show that this restriction naturally imposes a particular structure to both the encoder and to the decoder networks. This introduces prior information in the AEC without reducing the flexibility of the mixing model. Simulations with synthetic and real data indicate that the proposed strategy improves nonlinear HU.

\end{abstract}


\allowdisplaybreaks

\section{Introduction}

Hyperspectral Unmixing (HU) consists in unveiling the spectral signatures of pure materials, called \emph{endmembers} (EMs), and the proportions (also called \emph{abundances}) with which they appear at every pixel of a hyperspectral image (HI)~\cite{Dobigeon-2014-ID322}.
Although some HU methods assume the EMs spectra to be known \textit{a priori}~\cite{iordache2011sunsal,borsoi2018superpixels1_sparseU}, most applications require unsupervised algorithms, \cred{which estimate the EMs} from the HI~\cite{Borsoi_multiscaleVar_2018,qian2011unmixing_L12_NMF}.
\cred{The \emph{linear mixing model} (LMM) represents the reflectance of an observed pixel as}
a linear combination of the reflectance of the spectral signatures of the EMs, weighted by their corresponding abundance proportion.
However, the LMM fails to account for nonlinear interactions between different materials commonly seen in real scenes \cred{due to complex radiation scattering among several EMs}~\cite{Dobigeon-2014-ID322}. %

\cred{HU strategies considering  nonlinear mixtures}
can be generally divided into model-based and model-free methods. Model-based nonlinear HU assumes that the mixing process is known \textit{a priori}.
Common examples include algorithms based on, e.g., the bilinear mixing model (BLMM) or the post-nonlinear mixing model (PNMM)~\cite{Dobigeon-2014-ID322}. However, the mixing mechanisms can be complex in practice and an appropriate model is rarely available. This motivated the consideration of model-free nonlinear HU, which employs more flexible nonlinear mixing models which are able to represent the mixing process in practice and can be learned directly from the observed HI~\cite{Dobigeon-2014-ID322, Chen-2013-ID321}. Examples include the use of graph-based approximate geodesic distances~\cite{Heylen:2011kc} and kernel-based algorithms~\cite{Chen-2013-ID321,Imbiriba2016_tip}, the latter of which provides non-parametric function spaces that can represent arbitrary nonlinear mixtures.

Recently, the use of unsupervised neural networks (NNs) based on autoencoders (AECs) has become widespread in HU~\cite{guo2015autoencodersUnmixing,palsson2018autoencoderUnmixing_IEEEaccess,qu2018udas_autoencoderUnmixing}. 
AECs consist of encoder-decoder structured NNs originally devised for nonlinear dimensionality reduction. %
By associating the low-dimensional latent representation of the input pixels with the abundances, and the decoder structure of the network with the mixing model, HU can be performed by training the AEC on the observed HI~\cite{palsson2018autoencoderUnmixing_IEEEaccess}. The learned encoder is then applied to each image pixel to compute the abundances.
Several AEC-based strategies have been proposed for linear HU, using denoising autoencoders to reduce noise and outliers~\cite{guo2015autoencodersUnmixing,su2018autoencodersUnmixing,su2019deepAutoencoderUnmixing}, exploring sparsity constraints~\cite{qu2018udas_autoencoderUnmixing,ozkan2018endnet_autoencoderUnmixing}, using a near-orthogonality prior over the abundances~\cite{dou2020AEC_SU_hyperLapDDriven,dou2020dualB_AEC_nlSU} and convolutional architectures for spectral-spatial data processing~\cite{palsson2020convolutionalAEC_SU}, or using AECs as generative models to account for the spectral variability of the EMs~\cite{borsoi2019deepGun,borsoi2019EMlibManInterpVAE}.

More recently, AEC architectures have also shown promising performance in nonlinear HU, leading to algorithms that are unsupervised and model-free. For instance, in~\cite{wang2019AECnlin} a decoder network was proposed as the composition of an \cred{EM} matrix and nonlinear NN layers to learn post-nonlinear mixtures.
In~\cite{dou2020dualB_AEC_nlSU}, an AEC was proposed to account for bilinear mixtures by representing the abundances as the Hadamard product of two NN-generated estimates. However, the connection to the bilinear mixing model is not clear in this architecture.
In~\cite{zhao2019AECnlin} another architecture was presented by considering a decoder composed of a sum of a linear transformation and a multilayer NN \cred{to account} for other types of nonlinearity in the mixture.

Despite achieving good performance, existing AEC-based nonlinear HU algorithms fail to properly explore the fact that the encoder should invert the mixing process. This may reduce their robustness, especially when nonlinear NNs with many degrees of freedom are considered. 
In this paper, we propose a model-based AEC for nonlinear HU by considering the mixing model a nonlinear fluctuation over a linear mixture. Differently from previous works, we show that this restriction naturally imposes a particular structure to both the encoder and to the decoder networks. This introduces important prior information into the AEC without reducing the flexibility of the mixing model. Simulations with synthetic and real data indicate that the proposed strategy significantly improves the performance of nonlinear HU when compared to other state-of-the-art algorithms.

\vspace{-0.3cm}
\section{Problem formulation}
\vspace{-0.05cm}

The LMM assumes that each $L$-band pixel $\by_n\in\amsmathbb{R}^L$, $n=1,\ldots, N$, of an $N$-pixel HI, can be modeled as~\cite{Dobigeon-2014-ID322}:
\begin{align} \label{eq:LMM}
    &\by_n = \bM \ba_n + \be_n \,, %
    \,\,\,\, \text{s. t. }\,\,\,\, \ba_n\in\mathcal{S}^1
    \,, %
\end{align}
where $\bM \in \amsmathbb{R}^{L\times P}$ is a matrix whose columns are the $P$ endmember spectral signatures $\bm_k$, $\ba_n$ is the abundance vector, $\cp{S}^1=\{\bx\in\amsmathbb{R}^P:\bx\geq\cb{0},\,\cb{1}^\top\bx=1\}$ denotes the unity simplex, and $\be_n$ is an additive noise term.

Despite its popularity, the LMM fails to represent nonlinear interactions between the different materials in the scene which are commonly observed in real HIs~\cite{Dobigeon-2014-ID322}. This requires nonlinear mixing models, which can generally be represented as
\begin{align} \label{eq:NLMM}
    \by_n = \boldsymbol{f}(\bM,\ba_n) + \be_n \,, %
    \,\,\,\, \text{s. t. }\,\,\,\,
    \ba_n\in\mathcal{S}^1
    \,,%
\end{align}
in which $\boldsymbol{f}:\amsmathbb{R}^{L\times P}\times\amsmathbb{R}^{P}\to\amsmathbb{R}^L$ stands for different models for the interactions between the materials. Prominent examples include the BLMM (for macroscopic interactions), the PNMM (e.g., nonlinearities occurring between the scene and the sensor), and Hapke (for intimate mixtures) \mbox{models~\cite{Dobigeon-2014-ID322}.}

Since it can be difficult to specify a precise model for $\boldsymbol{f}$ in advance of HU, the use of nonparametric approaches based on, e.g., kernel machines, has received a lot of attention~\cite{Chen-2013-ID321,Imbiriba2016_tip,borsoi2020BMUAN}. Kernel methods have flexibility to model arbitrary nonlinear mixtures by learning $\boldsymbol{f}$ directly from the data. In this framework, a particularly interesting approach consists in assuming that the mixing process can be well represented as a nonlinear fluctuation $\boldsymbol{\psi}$ over the LMM~\cite{Chen-2013-ID321}:
\begin{align} \label{eq:NLMM2}
    &\by_n = \bM \ba_n + \boldsymbol{\psi}(\bM,\ba_n) + \be_n, %
    \,\,\,\, \text{s. t. }\,\,\,\,
    \ba_n\in\mathcal{S}^1 \,,
\end{align}
While the connection between \eqref{eq:NLMM} and \eqref{eq:NLMM2} may not be straightforward, model \eqref{eq:NLMM2} allows an easier control of the degree of nonlinearity in the model by penalizing the contribution of $\boldsymbol{\psi}$ during the HU process. In~\cite{Chen-2013-ID321}, an HU methodology was proposed using model~\eqref{eq:NLMM2} by constraining $\boldsymbol{\psi}$ to belong to a Reproducing Kernel Hilbert Space, what allowed an efficient solution as a convex optimization problem (i.e., a least-squares support vector regression problem). However, this methodology is purely supervised, and can not estimate the endmembers directly from the HI.

\vspace{-0.35cm}
\subsection{AEC-based unsupervised nonlinear HU}

More recently, unsupervised approaches based on AECs have been proposed for nonlinear HU. Such approaches differ from the AEC strategies used for linear HU in the sense that the decoder, which is based on the mixing model, must be designed to incorporate the nonlinearity seen in, e.g.,~\eqref{eq:NLMM} or~\eqref{eq:NLMM2}. For instance, in \cite{wang2019AECnlin} a network was proposed based on the post nonlinear model, where the decoder NN is formed by a composition of the endmember matrix and nonlinear layers. In~\cite{zhao2019AECnlin}, a similar idea was used based on model~\eqref{eq:NLMM2}, where the decoder NN was a linear combination of a linear layer (representing the EMs) and a nonlinear NN (representing the nonlinear interactions).
However, both these works use generic nonlinear network architectures for the encoder, which do not by themselves guarantee a proper inversion of the nonlinear HU process.
Being HU an inference problem, the objective is to retrieve both the abundances and EMs by fitting the model to a given HI. This contrasts with typical machine learning applications, where training and test sets are required to be distinct. Hence, the importance of structured models that guarantee the physical interpretability of the retrieved latent representations.

\vspace{-0.3cm}
\subsection{Importance of the encoder for HU}
\label{sec:encoder_importance}

Although some works have explored the influence of the encoder, its potential to improve the quality and robustness of AEC-based HU has not been fully utilized. For instance, early architectures considered tied weights for the encoder and decoder~\cite{guo2015autoencodersUnmixing}, which severely limited the performance. More recent works use almost exclusively untied weights~\cite{qu2018udas_autoencoderUnmixing}, using more flexible NNs as the encoder.
Denoising AECs have also been included as the first layers of the decoder in order to act as a pre-processing step before HU and improve the results~\cite{guo2015autoencodersUnmixing,su2018autoencodersUnmixing}. However, these works do not fully explore the facts that while the decoder should match the mixing process, the encoder should be constrained to represent its inverse. In the next section, we will leverage this knowledge to \cred{design the proposed} \mbox{model-based AEC.}

\begin{figure}
    \centering
    \includegraphics[width=1\linewidth]{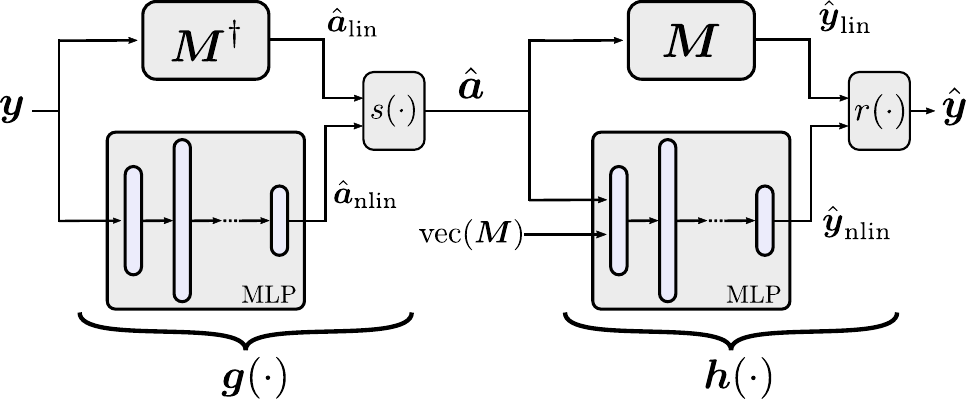}
    \vspace{-0.55cm}
    \caption{Proposed model-based autoencoder solution.}
    \label{fig:ours_diagram}
    \vspace{-0.4cm}
\end{figure}

\vspace{-0.2cm}
\section{Proposed solution}

In this work, we propose to leverage knowledge about the physical mixing process to derive an AEC architecture which is better able to represent \cred{both} the nonlinear mixing process \cred{and its inverse to perform HU in real world applications}. %
Our contributions in this step are twofold, related to the design of both the encoder and the decoder. An illustrative depiction is shown in Fig.~\ref{fig:ours_diagram}.

Inspired by~\eqref{eq:NLMM2}, we consider an AEC decoder design with an additive linear and nonlinear part which is similar to the one used in~\cite{zhao2019AECnlin}. The NN weights for the linear part of the decoder are directly associated to the EM signatures $\bM$. Moreover, we introduce both the abundances and the endmember signatures as inputs to the nonlinear part of the decoder. Therefore, by employing a NN architecture with high representation capacity, we can model arbitrary mixtures. This leads to a decoder $\bh:\amsmathbb{R}^P\to\amsmathbb{R}^L$ of the form:
\begin{align}
    \bh(\ba_n) = r\big(\bM\ba_n + \bomega_{D}(\ba_n,\bM;\bW_{\!D}) \big)\,,
    \label{eq:mdl_decoder}
\end{align}
where $\bomega_{D}$ is a nonlinear function in the decoder (e.g., an MLP) with parameters $\bW_{\!D}$ representing the nonlinear part of the mixing process, and $r:\amsmathbb{R}^L\to\amsmathbb{R}^L_+$ is a function that projects the decoder results to the nonnegative orthant.
Note that unlike in multi-kernel based methods such as~\cite{Chen-2013-ID321}, which consider nonlinearities in the mixing process to be only a function of $\bM$, $\bomega_{D}$ is a function of both the abundances and the EM signatures, which makes it more general.

The second part of the proposed design concerns the encoder, which we denote by $\bg:\amsmathbb{R}^L\to\amsmathbb{R}^P$. This has received far less attention from previous works, as discussed in Section~\ref{sec:encoder_importance}. Although most works adopt general linear or nonlinear NNs to design the encoder function, a better design can be obtained if we consider that the encoder should in principle be close to $\bh^{-1}$. For the case of linear HU (i.e., in which $\bomega_{D}\equiv\cb{0}$), this translates directly into a special kind of tied architecture, in which the encoder becomes $\bg\approx\bM^\dagger$~\cite{qu2018udas_autoencoderUnmixing}, where $\dagger$ denotes the pseudoinverse operator. Note that unlike previous works that consider tied architectures with $\bg\approx\bM^\top$~\cite{guo2015autoencodersUnmixing}, the architecture we mentioned is tied by the pseudoinverse, which is physically more reasonable. This has been used to motivate untied architectures, in which the encoder is left unconstrained~\cite{qu2018udas_autoencoderUnmixing}. However, this important knowledge has not been further employed to design or constrain neither the linear nor the nonlinear AEC encoder architecture due to the difficulty it introduces. %
Nonetheless, we can leverage this idea for nonlinear HU to make the method more principled and \cred{robust, %
while proposing a tractable training procedure.}

We propose to consider an encoder of the following form:
\begin{align}
    \bg(\by_n) & = s\big(\cred{\diag(\balpha)} \bM^{\dagger}\by_n + \bomega_{E}(\by_n;\bW_{\!E})\big) \,,
    \label{eq:mdl_encoder}
\end{align}
where $\bomega_{E}$ is a function parametrized on $\bW_{\!E}$ representing the nonlinear part of the encoder, and $s:\amsmathbb{R}^P\to\cp{S}^1$ is a function which maps the combinations from the linear and nonlinear abundance branch estimates to the unity simplex $\cp{S}^1=\{\bx\in\amsmathbb{R}^P:\bx\geq\cb{0},\cb{1}^\top\bx=1\}$ to ensure that the estimated abundances are physically meaningful. 
Parameter $\cred{\balpha\in\amsmathbb{R}^{P}_+}$ balances the contributions of the linear and nonlinear parts of the encoder, and is also a learnable parameter.
\cred{Note that the nonlinear parts of the encoder and decoder are closely related, in that to achieve small reconstruction errors the contributions of $\bomega_E$ and $\bomega_D$ must be similar.}

To illustrate this, suppose that a pixel $\by_n$ is generated from the decoder model with abundances $\ba_n$, that $r$ and $s$ are the identity function, $\cred{\balpha=\cb{1}}$, that $\bomega_D$ and $\bomega_E$ belong to normed function spaces. %
Then, to accurately reconstruct the abundances, we need $\|\ba_n - \bg(\bh(\ba_n))\|$ to be small. Using the reverse triangle inequality, this can generally \mbox{be written as:}
\begin{align}
    & \big\|\ba_n \!- \bg(\bh(\ba_n))\big\|
    \nonumber \\
    ={} & \big\|\ba_n \!-\! \bM^{\dagger}(\bM\ba_n \! + \! \bomega_D(\ba_n,\bM;\bW_{\!D}))  -  \bomega_E(\bh(\ba_n);\bW_{\!E})\big\|
    \nonumber \\
    ={} & \big\|-\bM^{\dagger}\bomega_D(\ba_n,\bM;\bW_{\!D}) - \bomega_E(\bh(\ba_n);\bW_{\!E})\big\|
    \nonumber \\
    \geq{} & \Big|\big\|\bomega_E(\bh(\ba_n);\bW_{\!E})\big\| - \big\|\bM^{\dagger}\bomega_D(\ba_n,\bM;\bW_{\!D})\big\| \Big|\,,
\end{align}
\cred{in which}
\begin{align}
    \big\|\bM^{\dagger}\bomega_D(\ba_n,\bM;\bW_{\!D})\big\| \leq \big\|\bM^{\dagger}\big\|\big\|\bomega_D(\ba_n,\bM;\bW_{\!D})\big\| \,.
\end{align}
\cred{This implies that} if $\|\bomega_D\|$ is small (small amounts of nonlinearity), we must necessarily have \cred{$\|\bomega_E\|$} small to obtain a small abundance reconstruction error. \cred{More generally, good abundance reconstruction requires the contributions of $\bomega_E$ and $\bM^{\dagger}\bomega_D$ to be similar for any amount of nonlinearity.} Therefore, if the norms of $\bomega_D$ and $\bomega_E$ are bounded above by the norm of the NN weights, we can account for different degrees of nonlinearity in a principled way by appropriately regularizing $\bW_{\!E}$ and $\bW_{\!D}$.
We call the proposed method MAC-U (Model-based AutoenCoder for \mbox{hyperspectral Unmixing).}

\subsection{Cost function}

The training process now consists in determining the proposed model-based encoder and decoder networks, $\bg$ and $\bh$, based on the set of available image pixels $\by_n$, $n=1,\ldots,N$. \cred{Note that, HU being an inference problem, we only have a single dataset (i.e., one HI) with $N$ pixels to both learn the model parameters and perform inference of the abundances using the AEC (this is in contrast to other traditional AECs that are trained and tested on separate datasets). Thus, the NN parameters must be learned using a single HI.}
A particular difficulty in learning the model-based AEC described in~\eqref{eq:mdl_decoder} and~\eqref{eq:mdl_encoder} is that it involves both $\bM$ and its pseuvoinverse $\bM^{\dagger}$. In order to obtain a more tractable architecture, let us rewrite~\eqref{eq:mdl_encoder} equivalently as
\begin{align}
    \!\!\!\!\bg(\by_n) & = s\big(\cred{\diag(\balpha)} \bQ\by_n + \bomega_{E}(\by_n;\bW_{\!E}) \big) \,\text{ s.t. } \, \bQ =\bM^{\dagger} 
    \label{eq:mdl_encoder2}
\end{align}
The constraint in~\eqref{eq:mdl_encoder2} is, however, difficult to enforce even for moderate numbers of endmembers $P$. Thus, we rewrite it as
\begin{align}
    \bQ &= \bM^{\dagger}
    = \big(\bM^\top\bM\big)^{-1} \bM^\top \Rightarrow \bM^\top\bM \bQ = \bM^\top \,.
    \label{eq:pseudoinv_constraint}
\end{align}
Denoting the network parameters by $\bW=\{\bW_{\!D},\bW_{\!E}\}$ and the training variables by $\bTheta=\{\bM,\bW,\bQ,\cred{\balpha}\}$, the cost function can be written as
\begin{align}
    \mathcal{L}(\bTheta) {}={} & \Ex_{\by\sim \cp{D}}\big\{\|\by - \bh(\bg(\by))\|^2 \big\} + \cp{R}_{\cal{W}}(\bW) + \cp{R}_{\cal{M}}(\bM)
    \nonumber\\
    & +\,\lambda_Q \big\|\bM^\top\bM \bQ - \bM^\top \big\|_F^2 \,,
    \label{eq:costf1}
\end{align}
where the expectation in the first term is taken with respect to the empirical distributions of the image pixels, supported at $\{\by_1,\ldots,\by_N\}$.
The constraint~\eqref{eq:pseudoinv_constraint} was introduced into the cost function in the form of an additive term. \cred{Function $\cp{R}_{\cal{W}}(\bW) \bigtriangleq \lambda_{\cal{W}} \big(\|\bW_{\!D}\|_{F}^2 + \|\bW_{\!E}\|_{F}^2 \big)$
is a regularization} which governs the nonlinear contributions to the encoder and to the decoder, \cred{while $\cp{R}_{\cal{M}}(\bM) \bigtriangleq \lambda_{\cp{M}}(\bm_k^\top\bm^{(0)}_{k})/(\|\bm_k\|\|\bm^{(0)}_{k}\|)$
constrains} the spectral angle between the updated endmembers and the initialization $\bM_0$. Parameters $\lambda_Q,\lambda_{\cal{W}},\lambda_{\cp{M}}\in\amsmathbb{R}_+$ balance the contributions of the regularizing terms in the cost function.

 \vspace{-0.15cm}
\subsection{Neural network architecture and cost function optimization}
\label{sec:nnet_architecture}

For the proposed MAC-U method, we considered the following NN architectures for the nonlinear part of the encoder and decoder blocks\cred{, adapted from the ones used in~\cite{zhao2019AECnlin}}. For $\bomega_E$, seven fully connected layers were used, with the leaky ReLU activation function and no bias term in the last layer. The layers contained $L$, $2L$, $L/2$, $L/4$, $4P$, $P$, and $P$ neurons, with non-integer values rounded up. For $\bomega_D$, five fully connected layers were used, with the leaky ReLU activation function and no bias term in the last layer. The layers contained $P(L+1)$, $PL$, $L$, $L$, and $L$ neurons. Function $r(\cdot)$ was implemented by a ReLU activation, and function $s(\cdot)$ by the normalized absolute value rectification mapping. 
$\mathcal{L}$ was minimized using the stochastic optimization method Adam~\cite{kingma2014adam}, with hyper-parameters set as: $\text{gradientDecayFactor} = 0.9$, $\text{squaredGradientDecayFactor} =0.95$, and $\text{miniBatchSize}=128$. Other hyper-parameters were left as the \cred{default} values in MATLAB$^\copyright$. Training was performed for at least one full epoch, and stopped when the relative change of $\mathcal{L}$ between two iterations was smaller than $0.01$.

\vspace{-0.15cm}
\section{Experiments}

This section illustrates the performance of the proposed method using simulations with both synthetic and real data. The proposed MAC-U method is compared to the fully constrained least squares (FCLS), to K-Hype~\cite{Chen-2013-ID321}, and to CDA-NL~\cite{halimi2016unmixingVariabilityNonlinearityMismodeling}. 
We also compare MAC-U with two other AEC-based methods. The first is a completely model-free AEC architecture (i.e., where neither the encoder nor the decoder have linear parts), which we call MF-AEC. The second is based on \cred{our implementation of} the architecture proposed in~\cite{zhao2019AECnlin}, which uses a linear model and nonlinear fluctuation only in the decoder (i.e., with a model-free encoder), which we call NF-AEC. Both MF-AEC and NF-AEC were implemented using the same framework and code as MAC-U, with the NN architectures described in Section~\ref{sec:nnet_architecture}.
In all experiments, EMs extracted from the observed HI using the VCA algorithm~\cite{Nascimento2005} were used for FCLS, K-Hype and CDA-NL, and as initialization for the different AEC strategies.
The performances of the methods were evaluated using the Root Means Squared Error (RMSE) between the estimated abundance maps ($\text{RMSE}_{\bA}$) and between the reconstructed images ($\text{RMSE}_{\bY}$). The RMSE is defined as $\text{RMSE}_{\bX} = \sqrt{\|\bX- \bX^*\|^2_F \,/\, N_{\bX}}$,
where $N_{\bX}$ denotes the number \mbox{of elements in $\bX$.}

\begin{table} [!ht]
\footnotesize
\centering
\caption{Quantitative results for data cubes DC1 and DC2.}
\vspace{-0.15cm}
\resizebox{\linewidth}{!}{%
\begin{tabular}{c||c|c|c|c||c}
\hline\hline
\multicolumn{6}{c}{DC1 data cube}  \\
\hline
& \multicolumn{2}{c|}{BLMM} & \multicolumn{2}{|c||}{PNMM}  \\
\hline
Method	&	$\text{RMSE}_{\!\bA}$	&	$\text{RMSE}_{\bY}$			&	$\text{RMSE}_{\!\bA}$	&	$\text{RMSE}_{\bY}$	 & \cred{Time}	\\
\hline	
FCLS	&0.2427 &0.0904 &0.1608 &0.0748  &0.85\\
K-Hype	&0.1859  &\textbf{0.0795}  &0.1523  &\textbf{0.0740} &7.54 \\ 
CDA-NL	&0.1965  &0.0808 &0.1628 &0.0745 &10.75\\
MAC-U (Prop.)	&\textbf{0.1050}  &0.0810 &\textbf{0.0795} &0.0747 &121.65\\
MF-AEC	&0.3460  &0.0891  &0.3693  &0.0793  &163.94 \\
NF-AEC	&0.2230  &0.0832 &0.1047 &0.0787 &75.34 \\
\hline\hline
\multicolumn{6}{c}{DC2 data cube}  \\
\hline
& \multicolumn{2}{c|}{BLMM} & \multicolumn{2}{|c||}{PNMM}  \\
\hline
Method	&	$\text{RMSE}_{\!\bA}$	&	$\text{RMSE}_{\bY}$			&	$\text{RMSE}_{\!\bA}$	&	$\text{RMSE}_{\bY}$	& \cred{Time}	\\
\hline				
FCLS	&0.1674 &0.1061 &0.0636 &0.0769 &0.85\\ 
K-Hype	&0.0992  &\textbf{0.0770} &0.0976 &\textbf{0.0759} &4.13\\ 
CDA-NL	&0.0910  &0.0780 &0.0616 &0.0765 &6.18 \\
MAC-U (Prop.) & \textbf{0.0864}  &0.0895 &\textbf{0.0512}  &0.0769 & 103.61\\
MF-AEC	& 0.3846  & 0.3943  & 0.1913  & 0.3287  &86.15  \\
NF-AEC	& 0.2337 & 0.1297  &0.1841  &0.0876  &80.10 \\
\hline\hline
\end{tabular}
}
\label{tab:quantitative_results}
\vspace{-0.15cm}
\end{table}

\paragraph*{\textbf{Synthetic data}}

Two synthetic datasets were considered, namely, Data Cube 1 (DC1), with $10^4$ pixels, and Data Cube 2 (DC2), with $2500$ pixels. Both DC1 and DC2 contained $P=3$ endmembers with 224 bands extracted from the USGS Spectral Library. The abundance maps were sampled from a Dirichlet distribution for DC1, and from a spatially correlated Gaussian random field for DC2.
The pixel reflectance values were generated using two nonlinear mixture models, namely, the BLMM:
$
	\by_n {}={} \bM\ba_n + \sum_{i=1}^{P-1} \sum_{j=i+1}^P  a_{n,i} a_{n,j} \bm_i \odot \bm_j + \be_n \,,
$
where $\odot$ is the Hadamard product, and the PNMM:
$
	\by_n {}={} (\bM\ba_n)^{0.7} + \be_n \,,
$
where the exponent is applied elementwise. Gaussian noise was added through $\be_n$ to result in a signal to noise ratio of 20~dB.
The parameters of the methods for each dataset were selected using a grid search within the ranges discussed in the original publications. For K-Hype, the parameter was selected among the values $\mu\in\{0.001,0.002,0.005,0.01,0.02,0.1,1\}$. For MAC-U and its variants, the parameters were selected by performing a grid search using the following values: $\lambda_Q,\lambda_\cp{W},\lambda_\cp{M}\in \{10^{-6}, 10^{-2}, 1\}$ and the learning rate $\gamma \in \{10^{-6}, 10^{-4}\}$.

\begin{figure}%
\centering
    \centering
    \includegraphics[height=0.95\linewidth, width=0.55\linewidth, angle =-90]{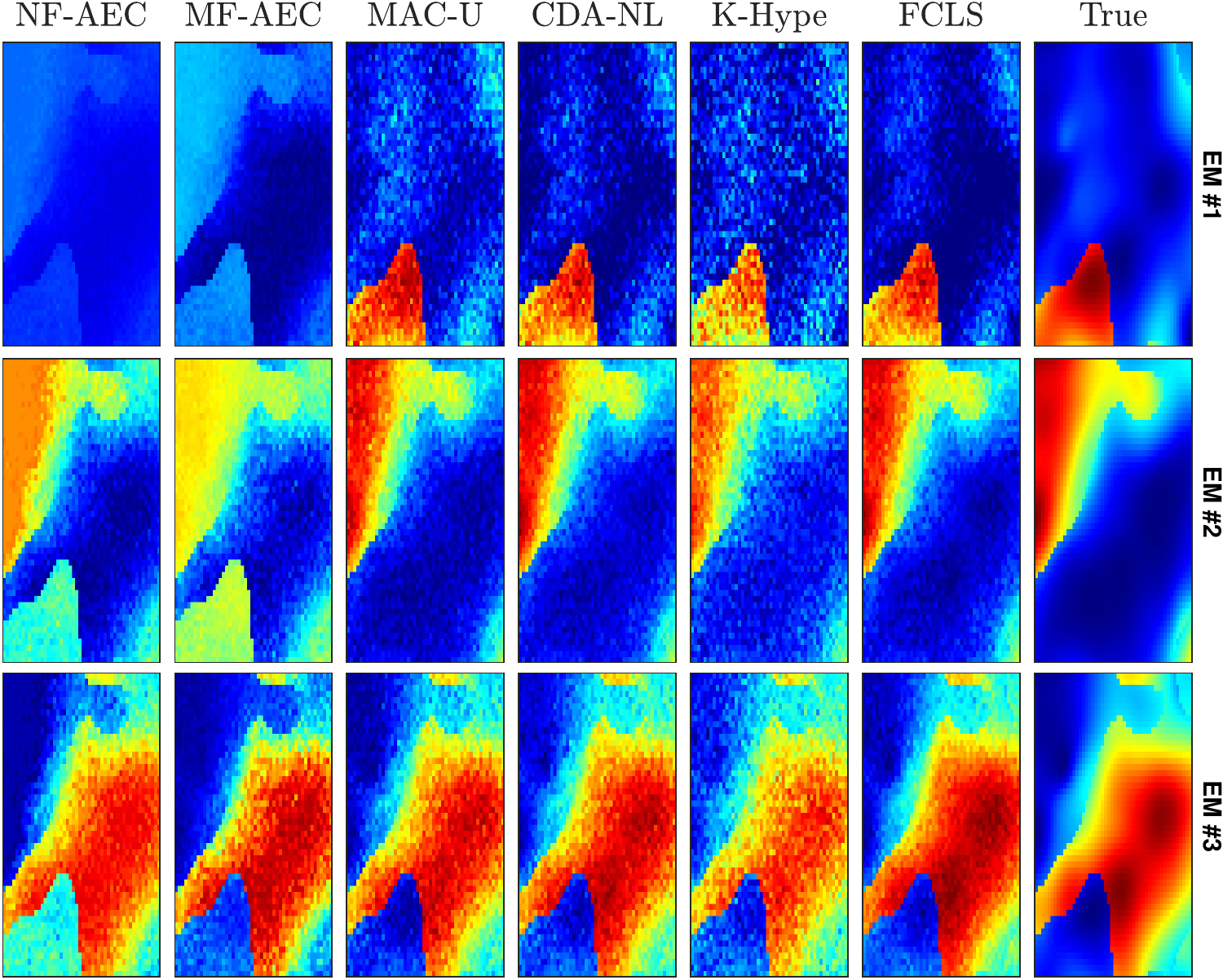}
\vspace{-0.15cm}
\caption{Abundance maps for DC2 for the PNMM model.}
\label{fig:dcs_abMaps}
\vspace{-0.4cm}
\end{figure}

The objective results are summarized in Table~\ref{tab:quantitative_results}. Fig.~\ref{fig:dcs_abMaps} presents the abundance maps only for DC2 with the PNMM due to space limitations. The proposed MAC-U outperformed the competing algorithms for all datacubes and nonlinearity models. CDA-NL also provided generally good results for DC1, while the performance of K-Hype varied according to the nonlinearity model. Moreover, the model-free architecture MF-AEC did not perform well, while NF-AEC (where the model is only enforced in the decoder) presented intermediate results. The proposed model-based architecture significantly improved the unmixing results among the AEC-based solutions. The $\text{RMSE}_{\bY}$ values for MAC-U were slightly larger than those of K-Hype, but smaller when compared to the other AEC solutions. However, we note that $\text{RMSE}_{\bY}$ is not directly related to the abundance reconstruction and thus not a good metric to evaluate the unmixing performance, especially when flexible models are considered. A visual inspection of the abundance maps in Fig.~\ref{fig:dcs_abMaps} corroborates the objective results.
\cred{The average execution times of MAC-U, shown in Table~\ref{tab:quantitative_results}, were significantly larger than the ones of KHype and CDA-NL, but compatible to those of the other AEC-based strategies.}

\paragraph*{\textbf{Real data}}

\begin{figure}[htb]
    \centering
    \includegraphics[height=0.4\textwidth, angle=-90]{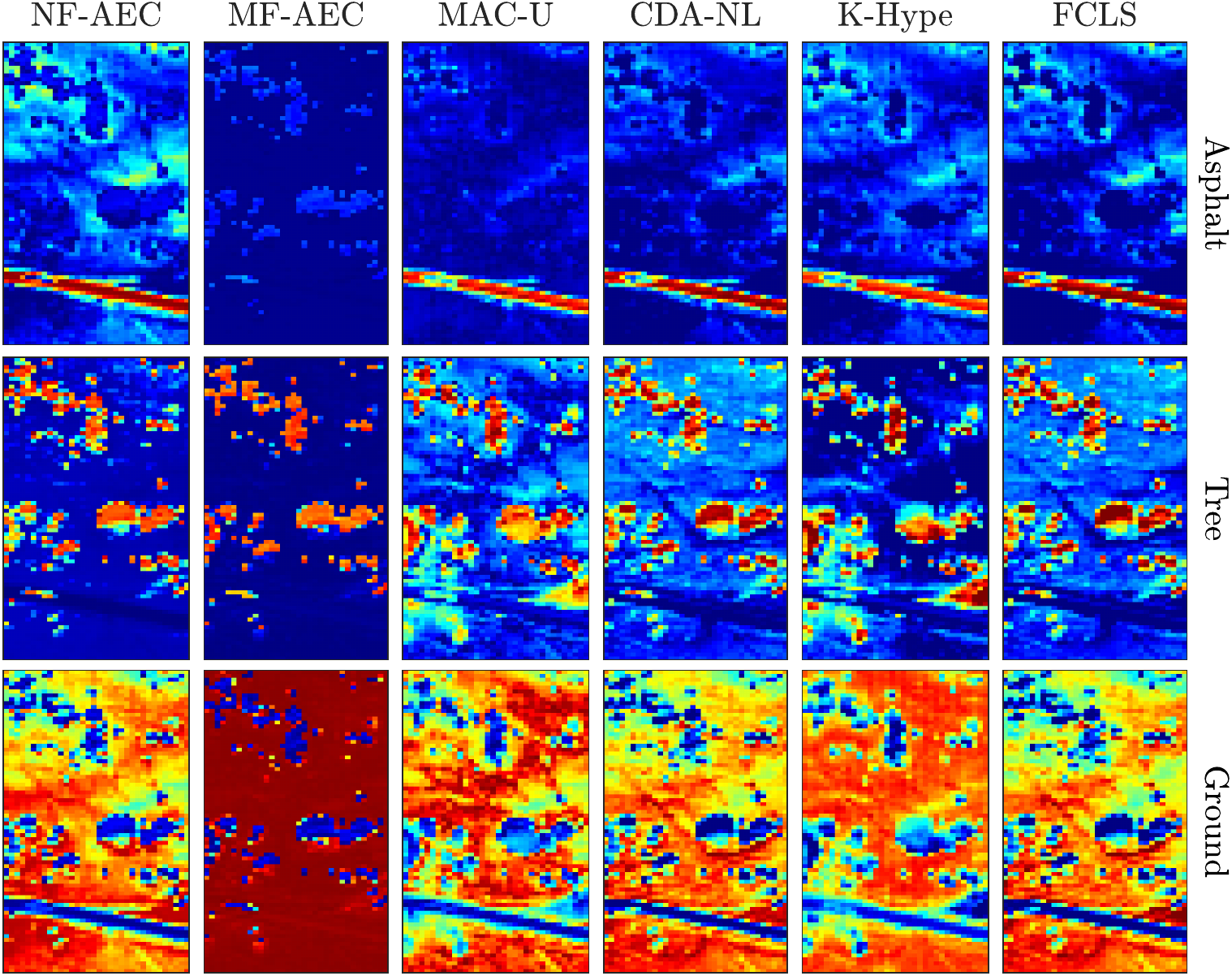}
    \vspace{-0.22cm}
    \caption{Abundance Maps of the Urban Scene.}
    \label{fig:urbanAbMaps}
\end{figure}

To evaluate the algorithms with real data we considered a subscene of the Urban HI~\cite{borsoi2020BMUAN}, with $L=162$ bands.
This subscene is known to contain 3 endmembers: Asphalt, Vegetation and Ground, where multiple scattering is expected to occur. Fig.~\ref{fig:urbanAbMaps} presents the abundance maps obtained with all the competing methods. By visual inspection one can notice that almost all methods present well defined and coherent abundance maps with exception of MF-AEC, which apparently merged the asphalt and ground EMs but presented a coherent map for the Tree abundance.
This behavior is expected due to the non-supervised and non-structured nature of this model. 
Regarding the remaining methods we can see an advantage of the proposed MAC-U algorithm for both the Asphalt and Ground endmembers while K-Hype seems to have provided a marginally better Tree abundance map, with energy concentrated in areas known to have vegetation. Nevertheless, the remaining methods, including the proposed MAC-U, also provide relatively accurate/coherent maps for the Tree endmember. \cred{The estimated linear scaling coefficients $\balpha = [1.0567,  0.9523 , 1.0402]^\top$ indicate strong contributions of the linear model. We highlight, however, that $\balpha$ values cannot directly measure the contribution of linear parcel of the model since the parameters of the nonlinear branch can grow to compensate posterior scaling parameters.}
The reconstruction errors of MAC-U, shown in Table~\ref{tab:realdata_results}, were comparable to those of FCLS and smaller than those of the other AEC-based architectures. However, the connection between small $\text{RMSE}_{\bY}$ and abundance reconstruction is not direct.

\vspace{-0.45cm}
\begin{table} [!ht]
\scriptsize
\centering
\caption{Quantitative results for the Urban Scene.}
\medskip
\vspace{-0.45cm}
\renewcommand{\arraystretch}{1.2}
\resizebox{\linewidth}{!}{%
\begin{tabular}{c||c|c|c|c|c|c}
\hline%
Method & FCLS & K-Hype & CDA-NL & MAC-U &  MF-AEC & NF-AEC \\
\hline				
$\text{RMSE}_{\bY}$	&0.02279 & \textbf{0.00610} & 0.00626 & 0.03756 & 0.12523 &0.04403 \\ \hline
\cred{Time} & 0.41 & 2.41 & 38.51 & 168.65 & 236.30 & 82.88 \\
\hline%
\end{tabular}
}
\label{tab:realdata_results}
\end{table}

\vspace{-0.45cm}
\section{Conclusions}
In this paper, a model-based autoencoder (AEC) network was proposed for nonlinear hyperspectral unmixing (HU). Considering the mixing model composed of a nonlinear fluctuation over a linear mixture, the proposed AEC can represent arbitrary nonlinear mixtures. Moreover, differently from previous approaches, the fact that the encoder should invert the mixing process was explicitly explored in this work. We showed that this restriction naturally imposes a particular structure to both the encoder and to the decoder networks, which explicitly makes use of the pseudoinverse of the endmember matrix. This introduced prior information into the AEC without reducing the flexibility of the mixing model. Simulations with synthetic and real data showed that the proposed strategy can improve the quality of nonlinear~HU.

\bibliographystyle{IEEEtran}
\bibliography{references,references2,references3}

\begin{thebibliography}{10}
\providecommand{\url}[1]{#1}
\csname url@samestyle\endcsname
\providecommand{\newblock}{\relax}
\providecommand{\bibinfo}[2]{#2}
\providecommand{\BIBentrySTDinterwordspacing}{\spaceskip=0pt\relax}
\providecommand{\BIBentryALTinterwordstretchfactor}{4}
\providecommand{\BIBentryALTinterwordspacing}{\spaceskip=\fontdimen2\font plus
\BIBentryALTinterwordstretchfactor\fontdimen3\font minus
  \fontdimen4\font\relax}
\providecommand{\BIBforeignlanguage}[2]{{%
\expandafter\ifx\csname l@#1\endcsname\relax
\typeout{** WARNING: IEEEtran.bst: No hyphenation pattern has been}%
\typeout{** loaded for the language `#1'. Using the pattern for}%
\typeout{** the default language instead.}%
\else
\language=\csname l@#1\endcsname
\fi
#2}}
\providecommand{\BIBdecl}{\relax}
\BIBdecl

\bibitem{Dobigeon-2014-ID322}
N.~Dobigeon, J.-Y. Tourneret, C.~Richard, J.~C.~M. Bermudez, S.~McLaughlin, and
  A.~O. Hero, ``Nonlinear unmixing of hyperspectral images: Models and
  algorithms,'' \emph{IEEE Signal Processing Magazine}, vol.~31, no.~1, pp.
  82--94, Jan 2014.

\bibitem{iordache2011sunsal}
M.-D. Iordache, J.~M. Bioucas-Dias, and A.~Plaza, ``Sparse unmixing of
  hyperspectral data,'' \emph{IEEE Transactions on Geoscience and Remote
  Sensing}, vol.~49, no.~6, pp. 2014--2039, 2011.

\bibitem{borsoi2018superpixels1_sparseU}
R.~A. {Borsoi}, T.~{Imbiriba}, J.~C.~M. {Bermudez}, and C.~{Richard}, ``A fast
  multiscale spatial regularization for sparse hyperspectral unmixing,''
  \emph{IEEE Geosci. Remote. Sens. Lett.}, vol.~16, no.~4, pp. 598--602, 2019.

\bibitem{Borsoi_multiscaleVar_2018}
R.~A. {Borsoi}, T.~{Imbiriba}, and J.~C. {Moreira Bermudez}, ``A data dependent
  multiscale model for hyperspectral unmixing with spectral variability,''
  \emph{IEEE Trans. on Image Proc.}, vol.~29, pp. 3638--3651, 2020.

\bibitem{qian2011unmixing_L12_NMF}
Y.~Qian, S.~Jia, J.~Zhou, and A.~Robles-Kelly, ``Hyperspectral unmixing via
  {L}$_{1/2}$ sparsity-constrained nonnegative matrix factorization,''
  \emph{IEEE Trans. Geosci. Remote Sens.}, vol.~49, no.~11, pp. 4282--4297,
  2011.

\bibitem{Chen-2013-ID321}
J.~Chen, C.~Richard, and P.~Honeine, ``Nonlinear unmixing of hyperspectral data
  based on a linear-mixture/nonlinear-fluctuation model,'' \emph{IEEE
  Transactions on Signal Processing}, vol.~61, pp. 480--492, Jan 2013.

\bibitem{Heylen:2011kc}
R.~Heylen, D.~Burazerovic, and P.~Scheunders, ``{Non-linear spectral unmixing
  by geodesic simplex volume maximization},'' \emph{IEEE Journal of Selected
  Topics in Signal Processing}, vol.~5, no.~3, pp. 534--542, 2011.

\bibitem{Imbiriba2016_tip}
T.~Imbiriba, J.~C.~M. Bermudez, C.~Richard, and J.-Y. Tourneret,
  ``Nonparametric detection of nonlinearly mixed pixels and endmember
  estimation in hyperspectral images,'' \emph{IEEE Transactions on Image
  Processing}, vol.~25, no.~3, pp. 1136--1151, March 2016.

\bibitem{guo2015autoencodersUnmixing}
R.~Guo, W.~Wang, and H.~Qi, ``Hyperspectral image unmixing using autoencoder
  cascade,'' in \emph{Proc. 7th Workshop on Hyperspectral Image and Signal
  Processing: Evolution in Remote Sensing (WHISPERS)}, Tokyo, Japan, June 2015,
  pp. 1--4.

\bibitem{palsson2018autoencoderUnmixing_IEEEaccess}
B.~Palsson, J.~Sigurdsson, J.~R. Sveinsson, and M.~O. Ulfarsson,
  ``Hyperspectral unmixing using a neural network autoencoder,'' \emph{IEEE
  Access}, vol.~6, pp. 25\,646--25\,656, 2018.

\bibitem{qu2018udas_autoencoderUnmixing}
Y.~Qu and H.~Qi, ``{uDAS}: An untied denoising autoencoder with sparsity for
  spectral unmixing,'' \emph{IEEE Transactions on Geoscience and Remote
  Sensing}, vol.~57, no.~3, pp. 1698--1712, March 2019.

\bibitem{su2018autoencodersUnmixing}
Y.~Su, A.~Marinoni, J.~Li, J.~Plaza, and P.~Gamba, ``Stacked nonnegative sparse
  autoencoders for robust hyperspectral unmixing,'' \emph{IEEE Geosci. and
  Remote Sens. Lett.}, vol.~15, no.~9, pp. 1427--1431, 2018.

\bibitem{su2019deepAutoencoderUnmixing}
Y.~Su, J.~Li, A.~Plaza, A.~Marinoni, P.~Gamba, and S.~Chakravortty, ``{DAEN}:
  Deep autoencoder networks for hyperspectral unmixing,'' \emph{IEEE Trans.
  Geosci. Remote Sens.}, vol.~57, no.~7, pp. 4309--4321, 2019.

\bibitem{ozkan2018endnet_autoencoderUnmixing}
S.~Ozkan, B.~Kaya, and G.~B. Akar, ``Endnet: Sparse autoencoder network for
  endmember extraction and hyperspectral unmixing,'' \emph{IEEE Transactions on
  Geoscience and Remote Sensing}, no.~99, pp. 1--15, 2018.

\bibitem{dou2020AEC_SU_hyperLapDDriven}
Z.~Dou, K.~Gao, X.~Zhang, H.~Wang, and J.~Wang, ``Hyperspectral unmixing using
  orthogonal sparse prior-based autoencoder with hyper-laplacian loss and
  data-driven outlier detection,'' \emph{IEEE Transactions on Geoscience and
  Remote Sensing}, 2020.

\bibitem{dou2020dualB_AEC_nlSU}
------, ``Blind hyperspectral unmixing using dual branch deep autoencoder with
  orthogonal sparse prior,'' in \emph{Proc. IEEE Int. Conf. on Acoustics,
  Speech and Signal Processing (ICASSP)}, Barcelona, Spain, 2020, pp.
  2428--2432.

\bibitem{palsson2020convolutionalAEC_SU}
B.~Palsson, M.~O. Ulfarsson, and J.~R. Sveinsson, ``Convolutional autoencoder
  for spectral-spatial hyperspectral unmixing,'' \emph{IEEE Transactions on
  Geoscience and Remote Sensing}, pp. 1--15, 2020.

\bibitem{borsoi2019deepGun}
R.~A. Borsoi, T.~Imbiriba, and J.~C.~M. Bermudez, ``Deep generative endmember
  modeling: {An} application to unsupervised spectral unmixing,'' \emph{IEEE
  Trans Comput Imaging}, vol.~6, pp. 374--384, 2019.

\bibitem{borsoi2019EMlibManInterpVAE}
R.~A. Borsoi, T.~Imbiriba, J.~C.~M. Bermudez, and C.~Richard, ``Deep generative
  models for library augmentation in multiple endmember spectral mixture
  analysis,'' \emph{IEEE Geoscience and Remote Sensing Letters (accepted)},
  2020.

\bibitem{wang2019AECnlin}
M.~Wang, M.~Zhao, J.~Chen, and S.~Rahardja, ``Nonlinear unmixing of
  hyperspectral data via deep autoencoder networks,'' \emph{IEEE Geoscience and
  Remote Sensing Letters}, vol.~16, no.~9, pp. 1467--1471, 2019.

\bibitem{zhao2019AECnlin}
M.~Zhao, M.~Wang, J.~Chen, and S.~Rahardja, ``Hyperspectral unmixing via deep
  autoencoder networks for a generalized linear-mixture/nonlinear-fluctuation
  model,'' \emph{arXiv preprint: 1904.13017}, 2019.

\bibitem{borsoi2020BMUAN}
R.~A. Borsoi, T.~Imbiriba, J.~C.~M. Bermudez, and C.~Richard, ``A blind
  multiscale spatial regularization framework for kernel-based spectral
  unmixing,'' \emph{IEEE Trans. on Image Proc.}, vol.~29, pp. 4965--4979, 2020.

\bibitem{kingma2014adam}
D.~P. Kingma and J.~Ba, ``Adam: A method for stochastic optimization,'' in
  \emph{Proc. International Conf. on Learning Representations (ICLR)}, 2015.

\bibitem{halimi2016unmixingVariabilityNonlinearityMismodeling}
A.~Halimi, P.~Honeine, and J.~M. Bioucas-Dias, ``Hyperspectral unmixing in
  presence of endmember variability, nonlinearity, or mismodeling effects,''
  \emph{IEEE Trans. on Image Proc.}, vol.~25, no.~10, pp. 4565--4579, 2016.

\bibitem{Nascimento2005}
J.~M.~P. Nascimento and J.~M. Bioucas-Dias, ``{Vertex Component Analysis}: A
  fast algorithm to unmix hyperspectral data,'' \emph{IEEE Transactions on
  Geoscience and Remote Sensing}, vol.~43, no.~4, pp. 898--910, 2005.

\end{thebibliography}

\end{document}